\documentclass[aps,pre,12pt,showpacs]{revtex4-1}
\usepackage{graphicx}
\usepackage{epstopdf}
\usepackage{amsmath}
\usepackage{amssymb}
\usepackage{default}
\usepackage{default}
\newcommand{\be}{\begin{equation}}
\newcommand{\bea}{\begin{eqnarray}}
\newcommand{\bc}{\begin{center}}            
\newcommand{\ee}{\end{equation}}
\newcommand{\eea}{\end{eqnarray}}
\newcommand{\ec}{\end{center}}

\newcommand{\baa}{\begin{eqnarray*}}
\newcommand{\eaa}{\end{eqnarray*}}
\begin{document}
\title{The many avatars of Curzon-Ahlborn efficiency}
 \author{Ramandeep S. Johal}
 \email{rsjohal@iisermohali.ac.in}
 \affiliation{ 
 Department of Physical Sciences, \\ 
 Indian Institute of Science Education and Research, Mohali \\
 Sector 81, S.A.S. Nagar, Manauli PO 140306, Punjab, India}
 \author{Arun M. Jayannavar}
 \email{jayan@iopb.res.in}
 \affiliation{Institute of Physics, \\ Sachivalaya Marg, Bhubaneswar 751 005, India}
\begin{abstract}
Efficiency at maximum power output of irreversible heat engines
has attracted a lot of interest in recent years.
We discuss the occurence of a particularly simple
and elegant formula for this efficiency in various different
models.  The so-called Curzon-Ahlborn efficiency is given by the 
square-root formula: $1-\sqrt{T_c/T_h}$, where $T_c$
and $T_h$ are the cold and hot reservoir temperatures.
\end{abstract}

 % \begin{abstract}
% a 
% \par\noindent
% Keywords: Power plants; Heat Engines; Thermal efficiency; Inference
% \end{abstract}
%\pacs{05.70.-a, 05.70.Ce, 05.70.Ln}

\maketitle
\section{Introduction}
Industrial revolution fueled our efforts to understand 
the working of thermal machines to improve their performance. 
This gave rise to the science of thermodynamics  
which identifies the permissible 
limits---put by physical laws---on our energy conversion devices. 
Carnot's insight on the maximum efficiency of a
heat engine operating between two heat reservoirs
at temperatures $T_{h}$ (hot) and $T_{c}$ (cold), yielded the universal
expression:
\be
\eta_C = 1- \frac{T_c}{T_h},
\label{ec}
\ee
which involves only the ratio of the temperatures, 
and is independent
of the details of the working medium. However, Carnot bound
is not realistic for actual heat engines and power plants.
Naturally, real machines operate under irreversibilities
caused by various factors, like finite rates of energy and matter flows, 
internal friction, heat leakage and so on, 
unlike the idealized quasi-static processes of textbook models.
Further, real engines are required to produce a finite power
output within a finite cycle time, whereas the power output of a
Carnot engine vanishes due to infeasibly large cycle times. 
In fact, the observed efficiencies are usually much less than 
the Carnot limit.
Thus the analysis of irreversible models with finite-rate
processes seems a desirable goal to pursue.

In recent years, there has been a great interest in extending  
thermodynamic models to justify the observed performance of industrial
power plants \cite{Curzon1975, Bejan1997, Esposito2010}.
An interesting quantity is the efficiency at maximum
power (EMP) of an irreversible model.
An elegant expression is often derived in some models, given as 
\be
\eta_{CA} = 1-\sqrt{\frac{T_c}{T_h}}.
\label{eca}
\ee

\begin{figure}[ht]
\includegraphics[width =9cm]{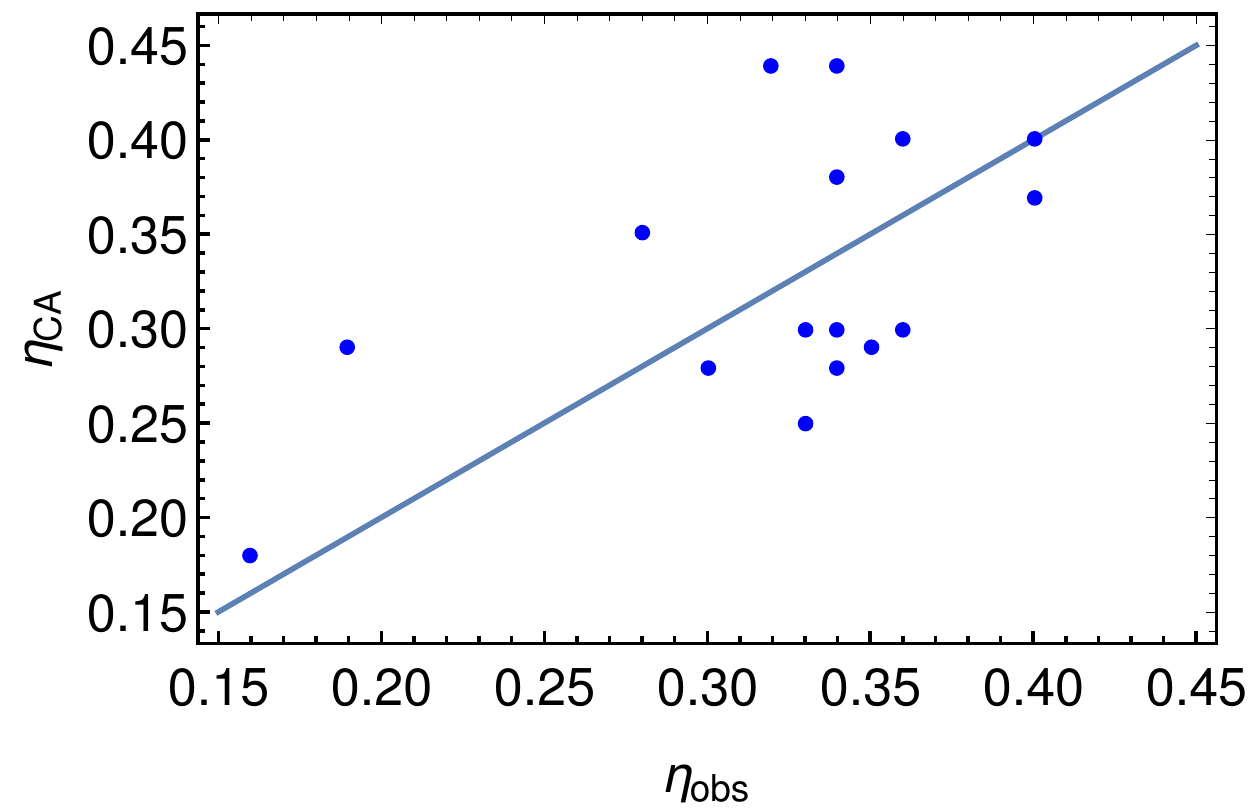}
 \caption{The observed efficiencies ($\eta_{\rm obs}$) 
 of power plants, plotted relative to the 
 CA-value for respective plants, as given in Ref. 
 \cite{Johalepj2017}.
 A data point lying on the diagonal  
 has the observed efficiency equal to CA-value. 
 For data points above the diagonal, $\eta_{\rm obs}$ is below
 the CA-value, while the converse is true for the points below the diagonal.}
\label{ecaf}
 \end{figure}
Due to historical priority \cite{Feidt2014}, the above expression is
also addressed as 
Reitlinger-Chambadal-Novikov efficiency (see Box I). However,  more recently 
it was rediscovered by Curzon and Ahlborn (CA) \cite{Curzon1975},
and in the physics literature,
it is popularly addressed as CA efficiency.
For small differences in the reservoir temperatures, this formula
yields the following behaviour:
\be
\eta_{CA} \sim \frac{\eta_C}{2} + \frac{\eta_{C}^{2}}{8} + \cdots
\ee
So to first approximation, CA efficiency is ``about'' one-half of the Carnot
value, which gives a reasonable estimate for the observed efficiencies
of many actual plants (see Fig. 1).

In fact, CA expression for the efficiency can be derived
by various different approaches and models. In this 
article, we highlight a few of the models which have
been studied in recent years.

\section{CA efficiency with finite-time Carnot engine}
The model proposed by Curzon and Ahlborn
 ascribes the only sources of irreversibilities 
to the finite rates of heat transfer across 
thermal resistances at both contacts 
with reservoirs (see Fig. 2). 
Further, it is assumed that the working medium remains
in internal equilibrium at temperature $T_1 < T_h (T_2> T_c) $
while exchanging heat with the hot (cold) reservoir. 
Assuming a Newtonian flow of heat,
the amount of heat flowing for time $t_1$ across
the hot contact is:
\be 
Q_h = \alpha (T_h - T_1) t_1.
\ee
Similarly, at the cold contact:
\be 
Q_c = \beta (T_2 - T_c) t_2,
\ee
where $\alpha, \beta$ are the heat transfer coefficients.
Thus, the work, $W= Q_h-Q_c$, is generated in time $t=t_1+t_2$.
Note that the time for the adiabatic steps can be 
taken to be negligible by assuming fast relaxation times
within the working medium. 

An important assumption in this model is that
the work generating compartment functions in
a reversible way. This means that the working
medium is in internal equilibrium, and this
is called the {\it endoreversible} assumption,
expressed as:
\be
\frac{Q_h}{T_1} =  \frac{Q_c}{T_2}.
\ee
Then, the average power
output of the cycle, defined as $P=W/t$, may be optimized over the 
contact times ($t_1, t_2$) \cite{Curzon1975},
or alternately, over the intermediate temperature variables ($T_1, T_2$).
It is found that EMP 
is given by the CA formula, which, interestingly, is independent
of the heat-transfer coefficients. Note that, obtaining
the CA result here depends on the choice of the phenomenological heat
transfer law. For instance, the choice of a different law gives a
different expression for EMP, and so it is not quite general.

\section{Linear Irreversible model}
There is another class of models, which do not consider 
a step-wise cycle of the working medium in a finite time.
These models consider the fluxes of entropy, matter and energy
in steady-state regime. 
 The theoretical framework is based on Onsager theory \cite{Joubook} 
 in which the fluxes ($J_k$) and their corresponding
 thermodynamic forces ($X_k$), or gradients, are identified. 
 Applied to a heat engine between two heat reservoirs, 
 suppose the engine performs work on the environment
against a fixed external force $F$. Then the work output
is $W = -F x$, where $x$ is the conjugate variable of the force. 
The power output is written as $\dot{W} = -F\dot{x}$,
where the dot represents the time derivative. 
Apart from the power flux, there is heat flux
arising due to temperature gradient. Thus
we can identify two flux-force pairs, as follows:
\bea
J_1 &=& \dot{x}, \quad X_1 = \frac{F}{T},\\
J_2 & = & \dot{Q}_h, \quad X_2 = \frac{\Delta T}{T^2},
\eea
where 
$T\approx T_h \approx T_c$, and $\Delta T= T_h-T_c$
is small compared to $T$. 
 Then the rate of entropy generation can be written
 in a bilinear form:
 \be
 \dot{S} = J_1 X_1 + J_2 X_2.
 \ee
 In the limit of small gradients, 
  each flux is assumed to be  a linear combination 
 of the available forces. Thus,
 \bea
 J_1 & = & L_{11} X_1 + L_{12} X_2, \\
 J_2 & = & L_{21} X_1 + L_{22} X_2,
 \eea 
 where the $L_{ij}$ are the phenomenological Onsager
 coefficients, satisfying  the reciprocity relation, $L_{12} = L_{21}$.
 Then, the power can be written as
\be
\dot{W} = -T J_1 X_1 = -T ( L_{11} X_1 + L_{12} X_2) X_1.
\ee
For given $X_2$, power may be optimized w.r.t $X_1$, which yields 
$X_{1}^{*} = -L_{12} X_2/2L_{11}$. Then the EMP, defined as 
$\eta^* = \dot{W} (X_{1}^{*})/\dot{Q}_{h} (X_{1}^{*})$,   is
evaluated to be:
\be
\eta^* = \frac{\Delta T}{2T} \frac{q^2}{2-q^2}.
\ee
Here, $q = L_{12}/\sqrt{L_{11} L_{22}}$,
denotes the strength of the coupling between the two fluxes,  
and satisfies $-1 \leq q \leq +1$.
At the tight-coupling condition $q^2 = 1$, 
we obtain the upper bound to EMP as $\Delta T/2T$, 
which is the half-Carnot value, or in other 
words, CA value is obtained to first approximation. 
 
Thus, within the framework of linear irreversible thermodynamics,
 valid for small values of forces or gradients, and
which assumes a linear dependence of fluxes  on 
the forces, the EMP
 is bounded from above by $\eta_{\rm C}/2$ \cite{Broeck2005}. 
Further, including conditions due to a certain left-right 
symmetry in the model, leads to the second order term of $\eta_{\rm C}^2/8$.
Thus, although the CA formula---in its  exact form---seems to be valid
under more specific conditions, there is a universality 
of EMP upto second order, based on more general 
thermodynamic considerations.

\begin{figure}[ht]
\includegraphics[width=8cm]{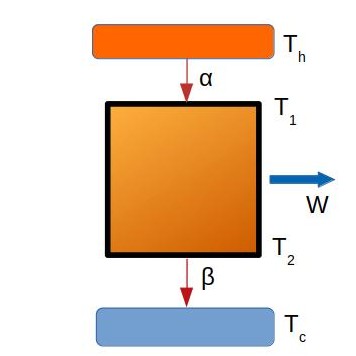}
\caption{Curzon Ahlborn model based on external irreversibilities, 
or the endoreversible assumption. The middle box represents the 
reversible, work-extracting compartment.}
\label{etfig2}
\end{figure}

\section{CA efficiency with a quasi-static model} 
\subsection{The case of complete information}
Consider a textbook example \cite{Callenbook1985} of 
two {\it finite} ideal-gas systems with a constant heat capacity $C$,
at initial temperatures $T_h$ and $T_c$, which serve  
as our heat source and the sink in their initial states, respectively. 
We also assume the availability of heat reservoirs at fixed
temperatures $T_h$ and $T_c$, by using which, the given systems
may be prepared or brought to their initial states. 

Now, the finite source and the sink are coupled to a {\it reversible} engine,
which extracts maximal work due to the available temperature gradient.
The total process will involve the two systems coming to mutual
equilibrium at a common final temperature. 
Suppose, at an intermediate stage, the source has acquired a 
temperature $T_1$ while the sink is at
 temperature $T_2$. The amount of heat
taken from the source by the engine is 
$Q_{\rm in} = C (T_{h}-T_{1})$,
while the heat rejected to the sink is
$Q_{\rm out} = C (T_{2}-T_{c})$.
The process being reversible, 
the total entropy change in the two systems is zero.
So, we have 
$\bigtriangleup S = C \ln {(T_{1}/T_{h})} + C\ln {(T_{2}/T_{c})} = 0$.
This yields the following relation between the final system temperatures 
\be
T_1 = \frac{T_h T_c}{T_2}.
\label{a}
\ee
The extracted work,  
$W= Q_{\rm in}-Q_{\rm out}$, is given by  
\be
W= C (T_{h}+T_{c}-T_{1}-T_{2}).
\label{b}
\ee
Using Eq. (\ref{a}), the efficiency $\eta= W/Q_{\rm in}$,
 can be written as:
\bea
\eta &=&  1 -  \frac{T_c}{T_1}, \label{et1}\\
     &=&  1 -  \frac{T_2}{T_h}. \label{et2}
\eea
We note that the work is maximum  when the final temperatures 
of the systems are equal: $T_1 = T_2 = \sqrt{T_hT_c}$, 
and so the efficiency of the  
total process is  $\eta_{\rm CA}$. 
Thus CA-efficiency emerges as the efficiency at maximum total 
work from finite ideal-gas systems. 

For the curious reader,
we wish to point out that by considering arbitrary thermodynamic 
systems  as our source and sink,
and possibly of different sizes,
a general expression for efficiency can be derived \cite{JohalRai2016}. 
It also leads to the $\eta_{\rm C}/2$ result  in the linear response 
regime, i.e. upto first
order in temperature differences. When the sizes of the two systems are similar, 
it leads to the universal 1/8 factor in the second order
term. Thus, we note that universality of efficiency, at
maximum power (finite-time models), and at maximum work
(quasi-static models), share similar features.
 
\subsection{Model with partial information} 
In the following, we introduce 
a completely different mechanism for arriving 
at CA efficiency.
We shall first discuss the set-up, of finite-sized source and sink,
from the point of view of a partial information and 
employ the methodology of inference (see Box II).

 Now, from Eqs. (\ref{a}) and (\ref{b}),
 the work performed by our engine can be rewritten as
\be
W(T_2) =  T_h +T_c -T_2  -\frac{T_h T_c}{T_2}.
\label{wt2}
\ee 
We let $C=1$, for simplicity. 
Clearly, a similar expression for work can be written in terms of $T_1$.
So, just from the expression for work, 
it is not apparent as to which system (source or sink) 
the value $T_2$ belongs to. We have to look
at the expression for the heat exchanged to assign
a label corresponding to a specific system (in the above case, 
$T_2$ refers to the temperature of the sink).  
Clearly, in the standard
thermodynamic model, the information conveyed by the symbol $T_2$, 
 consists of two parts: i) the individual
value of the temperature and ii) the label for the 
system to which it is assigned. 

Now, suppose that we do not extract the maximal total work from the 
systems, 
but abort the process  when 
the temperatures of one of the systems is $T$, and so, that of
the other is $T_h T_c/T$.
We imagine an observer who completes the cycle---bringing
the systems back to their initial states by putting them 
in contact with respective reservoirs. In the case
of an exact information, our observer makes no mistakes, and 
puts back the correct system with its respective reservoir.
But, suppose, there is a probability of making an error in 
this last step. Maybe, 
the observer is careless and does not read the labels properly,
or the labels may have got partially erased, and so on. Let 
us quantify the odds of this error with a probability, i.e a certain
number $0\le \gamma \le 1$.

Now, given the above state of partial information, 
the task ahead of the observer is to make an estimate of 
the work performed, the efficiency of the process and so on.
As mentioned above, the work expression does not reveal 
unambiguously the individual labels of the systems.
Thus given one temperature value as $T$, the work expression
will be written as:  
$W(T) =  T_h +T_c -T  -{T_h T_c}/{T}$.

To illustrate how the estimates may be based 
on our state of knowledge,  
 suppose the observer is completely ignorant of
 the exact labels for the systems, which means
 he makes the maximal error, and given a system, he 
 assigns equal probabilities that it may be the source
 or the sink, i.e., we assume $\gamma = 1/2$.
Then, the estimated temperature of {\it each} system can be given 
by the mean value over the two possible values:
\be
\overline{T} = \frac{1}{2} \left(T + \frac{T_hT_c}{T}\right).
\label{bart}
\ee
Now, we may estimate other
quantities relevant to the performance of the engine.
Our estimate of the heat absorbed by the engine is
\be
Q_{\rm in} = T_h - \overline{T}.
\label{qin}
\ee
Similarly,  our estimate for the heat rejected to the sink will be: 
\be
Q_{\rm out} = \overline{T} - T_c.
\label{qout}
\ee
The estimate for work, $W$, 
turns out to be the same, i.e.
equal to the actual work performed, 
showing that the work is not affected
by uncertainty in the labels of the final states of the systems. 

For brevity, let us normalize all temperatures w.r.t to 
the initial temperature of the source, and  define $\theta = T_c/T_h$
and $\tau = T/T_h$. So the estimate for work is  
\be
W(\tau) = 1+ \theta - \tau - \frac{\theta}{\tau}.
\label{wtth}
\ee
The estimate for the efficiency $\eta = W/Q_{\rm in}$ is then

\be
\eta_{{1}/{2}}(\tau) = \frac{2(\tau + \theta \tau - \tau^2-\theta)}{2 \tau-\tau^2-\theta}.
\label{etat}
\ee 
Now, if a given temperature  value, $\tau$, 
actually belongs to the source, then the efficiency
is: $\eta = 1-\theta/\tau$, whereas if this value
belongs to the sink, then $\eta = 1-\tau$ (see Eqs. (\ref{et1}) and (\ref{et2})). In the
case of maximal error, the efficiency is as in Eq. (\ref{etat}).

\begin{figure}[ht]
\includegraphics[width=8cm]{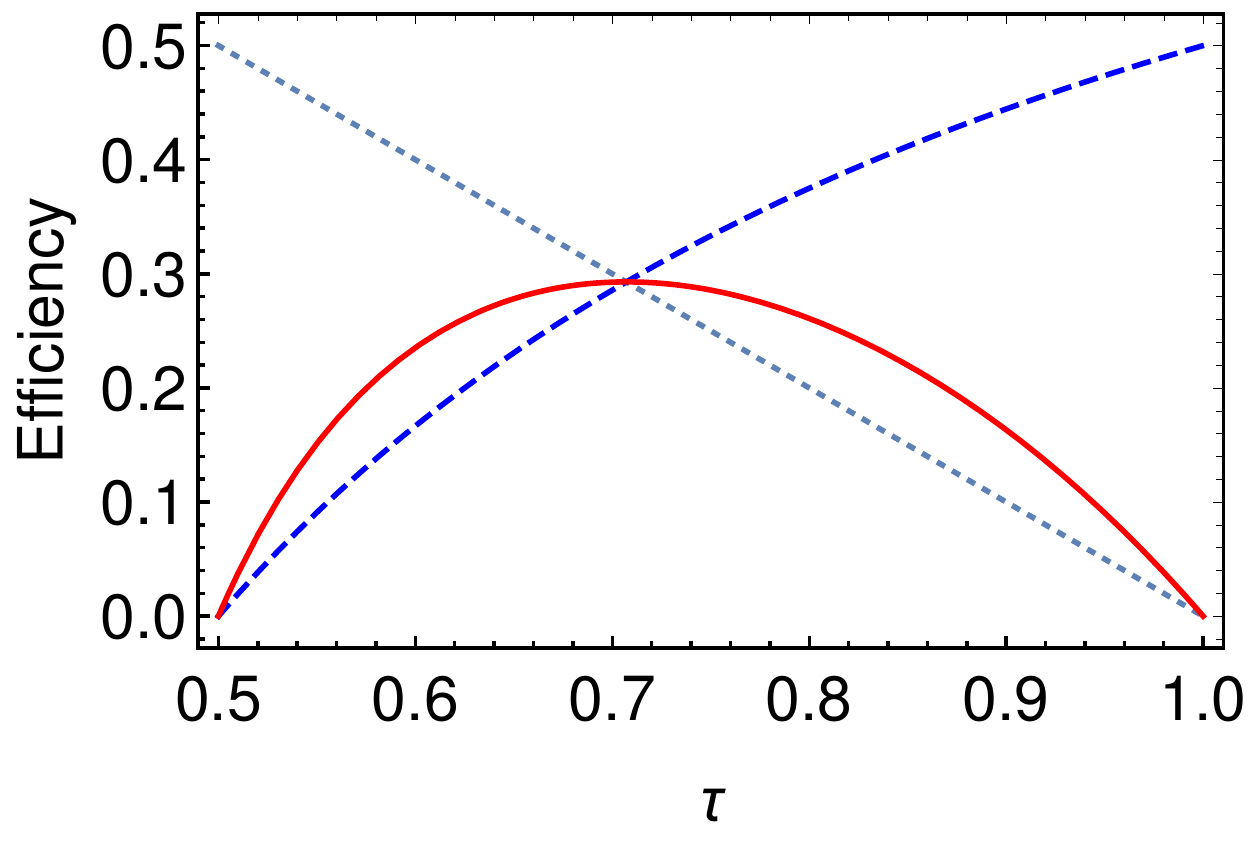}
\caption{Plots of efficiency for $\theta=0.5$. $\tau$ takes 
values in the range $[\theta,1]$. 
Dashed curve is $\eta = 1-\theta/\tau$, while the 
dotted curve is  $\eta = 1-\tau$. The solid curve
is the estimate as in Eq. (\ref{etat}), whose maximal value
is at $\tau = \sqrt{\theta}$ and is equal to CA value. Carnot limit 
is 0.5.
}
\label{etfig3}
\end{figure}
From Fig. 3, we see that in both   
cases of certainty about the labels (dotted and dashed curves),
 the maximum efficiency is the Carnot value ($1-\theta$). 
But,  an uncertainty in the correct labels reduces the 
maximal efficiency below the Carnot value, which 
 reduces to CA value in the case of
 maximal uncertainty.
So, here we have a mechanism which shows how inference under
partial information leads us to expect a lower value for the maximum
efficiency obtainable in a heat engine.
 Thus, the upper bound for the efficiency is directly 
related to our state of knowledge and CA 
efficiency emerges from an entirely different perspective \cite{Johal2015}.

Finally, we would like to illustrate the
important role played by Bayes' theorem (see Box III)
in the theory of inference. Let us consider
a two-level system with energy levels 0 and $a$, 
 denoted by symbols ($\downarrow$) and ($\uparrow$), respectively.
When we put this system in contact with a
heat reservoir at a certain temperature $T$, 
the equilibrium Boltzmann distribution is given by
$p(\uparrow \vert a) = 1/(1+\exp(a/k_B T))$, and 
$p(\downarrow \vert a) = 1- p(\uparrow \vert a)$.
These probabilities can be regarded as conditional
upon the given values of the energy gap $a$ and 
the temperature $T$. Now, suppose, we are ignorant
about the exact value of the gap parameter $a$. So, instead we
treat quantity $a$ as a random variable in the 
sense of subjective probability. 
We assign a prior distribution
$\Pi (a)$,
which for simplicity, may be taken as 
 uniform density over some interval, $[0,a_{\rm max}]$, 
i.e. $\Pi(a) = 1/a_{\rm max}$.
Note that this range is to be
based on some background information such as
the permissible values (real positive, in this case) and 
the maximal gap that may be obtained in the 
laboratory or under given physical conditions, 
and so on. Let us now make 
a measurement on the system to know whether it is 
in the ground or excited state. Based on the result
of the measurement,
we can use Bayes' theorem to update our probability for the gap as
follows:
\begin{equation}
p(a\vert \uparrow) d a = \frac{p(\uparrow \vert a ) \Pi (a) d a
}{\int_{0}^{a_{\rm max}} 
p(\uparrow \vert a ) \Pi (a) d a},
\label{Btheorem}
\end{equation}
which is a continuous-variable version, in contrast to 
the discrete case discussed in Box III.
In order to appreciate the product form of the probabilities in 
the above equation, note that $\int 
p(\uparrow \vert a ) \Pi (a) d a \equiv p(\uparrow)$.

Now, initially, we have assumed  all values of the gap 
to be equally likely (uniform prior).
Then, upon measurement, if we
find the system in ($\uparrow$) state, our revised
guess or the posterior probabilities should assign more weight
to smaller values of the gap, because these
are more likely to cause the system to absorb
energy from the reservoir and make it jump to ($\uparrow$) state.
Otherwise, if the system is found in ($\downarrow$) state, 
then our posterior guess would be that the larger values of the gap
are more likely. These inferences are illustrated in Fig. 4.
Application of Bayes' theorem in the above
fashion to quantum heat cycles, where  such 
a two-level system acts as a working medium,
leads to the estimates of the work per cycle which becomes optimal 
at CA-efficiency, under certain conditions. 
Remarkably, this feature holds for a
class of prior distributions \cite{Johal2010}.

\begin{figure}[ht]
 \includegraphics[width=8cm]{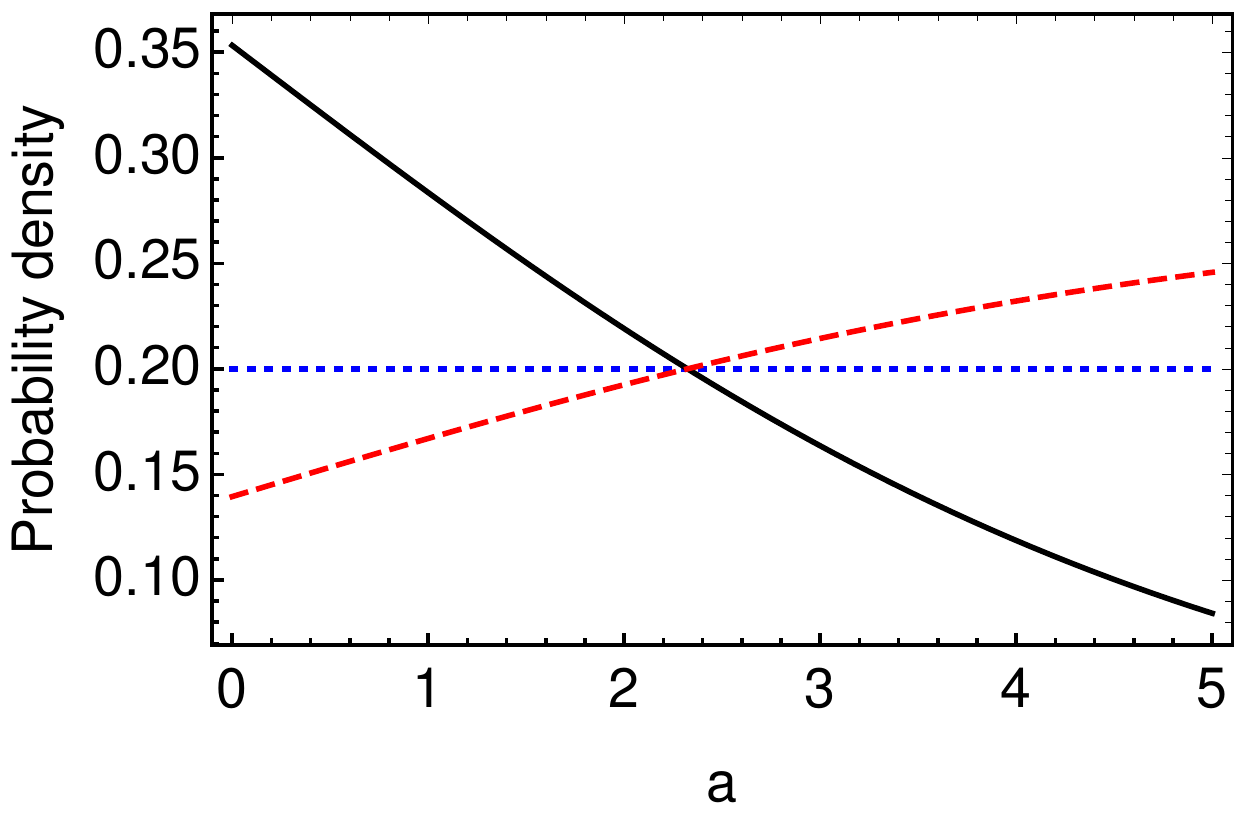}
 \caption{Probability density for the energy gap $a$ in 
 two-level quantum system.
 Uniform prior is shown as dotted horizontal line. If
 the measurement shows the system in ($\uparrow$) state, the posterior
 density is given by solid line, indicating that 
 the density becomes higher for smaller gap values.
 In case, the measurement shows ($\downarrow$) state, then 
 the posterior density becomes higher for larger gaps (dashed line). 
 Here, $a_{\rm max} = 5$, temperature $T=2.5$, and $k_B=1$.}
\end{figure}

\section{Summary}
Efficiency at maximum power output is an important 
figure of merit in the current research
on nonequilibrium thermodynamics. 
Due to the simplicity and elegance of the CA formula,
universal properties of the efficiency have 
been intensively studied in recent years. We have seen that
 CA value is obtained in a variety of models, 
ranging from finite time to quasi-static 
models of work extraction, including 
more abstract models of heat engines
with partial thermodynamic information. Although other 
expressions for efficiency are also derived
in different models, CA value
provides an important benchmark against
which these values may be compared,
as, for instance, 
in the observed efficiencies of real power plants.

\newpage
%\noindent
\fbox{%
    \parbox{14cm}{%
\subsection*{Box I: History of CA formula}
    The earliest known such model (1929) which derived the CA-efficiency,  
    is ascribed to Reitlinger \cite{Reitlinger}.
It involved a heat exchanger receiving heat from a finite hot stream
fed by a combustion process. An analogous model (1957) was applied to  a steam 
turbine by Chambadal \cite{Chambadal}. The considered heat exchanger 
in these models is effectively infinite. Novikov \cite{Novikov} considered 
the heat transfer process (1958) 
between a hot stream and a finite heat exchanger with a given heat conductance. 
There are two simple, but significant assumptions used in these models: 
i) a constant specific heat of the inlet hot stream and ii) the validity
of Newton's law for heat transfer. 
Further, there appears a floating temperature-variable  
in between the highest and the lowest values, such as the 
temperature of the exhaust warm stream, over which the output
power can be optimized. This yields an optimal value 
of the intermediate temperature which 
is usually found to be $\sqrt{T_h T_c}$, the geometric mean of $T_h$ and $T_c$.
This leads directly to the result that the 
efficiency at maximum power is the CA value.
The CA model (1975) \cite{Curzon1975}, 
which was apprarently posed as a classroom problem,
attracted the attention of the physics community and 
and it led to a new field 
termed as Finite-time Thermodynamics \cite{Andresen2011}.
In the engineering literature,
the analogous approach is called 
Entropy Generation Minimization \cite{Bejan1996, Bejan1997}.
\vskip 12pt
    }%
}

\newpage
\fbox{%
    \parbox{14cm}{%
\subsection*{Box II: Prediction vs. Inference}
It is important to appreciate
the significance of inference in science, in general.
Whereas prediction is concerned with the question of finding out the consequences when 
the causes are given,  
inference, on the other hand, deals with guessing the plausible causes, given the consequences. 
The questions of interest in the latter category, are also called inverse problems. For example,
a traditional question may be to guess the diffraction pattern, given the size and shape of the
aperture. But the corresponding question of inference would be that, given the pattern,
one is required to guess the possible shape of the aperture. Inference 
is a useful tool when only a limited or partial information is available.
Thus, another domain of inference is probability theory. 
A standard problem may be that 
given a certain proportion of black and white balls
in a closed bag, what is the probability that a white ball will show up in the next draw?
The problem of inference could be that the results of previous, finite number of draws are given
and one is required to guess the proportion of white/black balls in the bag. 
Usually, inference is based on plausible reasoning, which makes a rational
guess based on incomplete information, and arrives at the estimates for the   
uncertain quantities (see also Box III).  
}
}

\newpage

\fbox{%
    \parbox{15cm}{%
\subsection*{Box III: Bayes' Theorem}
In Bayesian approach to probability theory, all uncertainty is 
treated probabilistically. Thus, probability can be interpreted
in a subjective sense---as the degree of rational belief
of the observer based on his/her state of incomplete or 
partial knowledge about the system.
A central role is played in  this framework by the notion
of prior probabilities, which reflect the state of background
knowledge of the observer, before the data from observation 
is included in the analysis. Then, 
Bayes' theorem is a tool by which the 
initial (prior) guess of probabilities for certain events/hypotheses is 
updated whenever a new piece of information or data
is incorporated. To take a simple example, scientists may 
have some conjectures about the possibility
of life on a certain planet, based on previous data. 
Then, suppose that the presence of water is discovered on this 
planet. Now, since
water represents a distinct potential for life---as we know it, 
so the estimates of scientists about
the existence of life on that planet are expected
to change. Bayes' theorem can be a useful tool here,
if we are able to model its essential elements. 
More precisely, let $B$ represent the background 
knowledge before the discovery of water which 
is depicted as event $W$, and $L$ be the fact
that life exists on the planet. 
Then $p(W|B)$ and $p(L|B)$ are the {\it prior}
probabilities for the existence of water and life, respectively, 
on the planet. Similarly, $p(W|L,B)$
denotes the (conditional) probability for water, 
given that life exists there. 
 Then, we are interested in $p(L|W,B)$, 
the updated or {\it posterior} probability for the 
existence of life, after water is discovered. This 
can be obtained from the product law of probabilities, i.e.
\be
p(L|W,B) p(W|B) = p(W|L,B) p(L|B).
\label{bwl}
\ee
This is the essence of Bayes' theorem, which takes
into account how the new information updates our
previous beliefs and their probabilities.
Now, probabilities are numbers, which by definition,
lie between zero and unity. Thus, an assignment of unit value
implies certainty. In a specific scenario, we may 
assign $p(W|L,B)=1$, if we feel sure about the presence
of water, in case, life is already known to exist. Then, we can 
conclude from the above that $p(L|W,B) > p(L|B)$. Thus our 
or scientists' expectation
about life on the planet increases, after the discovery of water.
}
}

\newpage

\section*{Acknowledgements}
AMJ thanks Department of Science and Technology, India
for the grant of J.C. Bose National Fellowship.

%\bibliography{bibliofeynman}

%\bibliography{Means}


\end{document}